\documentclass[fleqn,usenatbib]{mnras}

\usepackage{newtxtext,newtxmath}
\usepackage[T1]{fontenc}
\usepackage{ae,aecompl}

\hypersetup{draft}
\usepackage{graphicx}
\usepackage{amsmath}
\usepackage{gensymb}
\usepackage{threeparttable}
\usepackage{booktabs,siunitx}
\usepackage{tabularx}
\usepackage{dcolumn}
\newcolumntype{d}[1]{D{.}{.}{#1}}

\title[Radio-GW prospects for A+ and beyond]{Radio Afterglows from Compact Binary Coalescences: Prospects for Next-Generation Telescopes}

\author[D. Dobie et al.]{Dougal Dobie,$^{1,2,3,4}$\thanks{E-mail: ddobie@swin.edu.au (DD)}
Tara Murphy$^{1,3}$
David L. Kaplan,$^{5}$
Kenta Hotokezaka,$^{6,7}$
\newauthor
Juan Pablo Bonilla Ataides,$^{1}$
Elizabeth K. Mahony,$^{2}$
Elaine M. Sadler$^{1,2}$
\\
$^{1}$Sydney Institute for Astronomy, School of Physics, University of Sydney, NSW 2006, Australia\\
$^{2}$ATNF, CSIRO Astronomy and Space Science, PO Box 76, Epping, NSW 1710, Australia\\
$^{3}$ARC Centre of Excellence for Gravitational Wave Discovery (OzGrav), Hawthorn, Victoria, Australia\\
$^{4}$Centre for Astrophysics and Supercomputing, Swinburne University of Technology, Hawthorn, Victoria, Australia\\
$^{5}$Department of Physics, University of Wisconsin-Milwaukee, Milwaukee, WI 53201, USA\\
$^{6}$Department of Astrophysical Sciences, Princeton University, 4 Ivy Lane, Princeton, NJ 08544, USA\\
$^{7}$Research Center for the Early Universe, Graduate School of Science, University of Tokyo, Bunkyo-ku, Tokyo 113-0033, Japan\\
}

\date{Accepted XXX. Received YYY; in original form ZZZ}

\pubyear{2021}
\begin{document}
\label{firstpage}
\pagerange{\pageref{firstpage}--\pageref{lastpage}}
\maketitle

\begin{abstract}
The detection of gravitational waves from a neutron star merger, GW170817, marked the dawn of a new era in time-domain astronomy. Monitoring of the radio emission produced by the merger, including high-resolution radio imaging, enabled measurements of merger properties including the energetics and inclination angle. In this work we compare the capabilities of current and future gravitational wave facilities to the sensitivity of radio facilities to quantify the prospects for detecting the radio afterglows of gravitational wave events. We consider three observing strategies to identify future mergers -- widefield follow-up, targeting galaxies within the merger localisation and deep monitoring of known counterparts. We find that while planned radio facilities like the Square Kilometre Array will be capable of detecting mergers at gigaparsec distances, no facilities are sufficiently sensitive to detect mergers at the range of proposed third-generation gravitational wave detectors that would operate starting in the 2030s. \end{abstract}


\section{Introduction}
The 2015 discovery of gravitational waves produced by a binary black hole merger by the Laser Interferometer Gravitational Wave Observatory (LIGO) kickstarted a new era of astronomy \citep{2016PhRvL.116f1102A}. Despite comprehensive follow-up observations carried out using telescopes across the electromagnetic spectrum, no electromagnetic counterpart was detected \citep{2016ApJS..225....8A,2016ApJ...826L..13A}, nor were any coincident neutrinos \citep{2016PhRvD..93l2010A}. Almost two years later LIGO/Virgo detected gravitational waves from a neutron star merger \citep{2017PhRvL.119p1101A}, which was accompanied by the contemporaneous detection of a short gamma-ray burst \citep[][]{2017ApJ...848L..13A,2017ApJ...848L..14G}. Approximately 10 hours later a candidate optical counterpart was discovered \citep{2017Sci...358.1556C,2017ApJ...848L..16S,2017ApJ...848L..24V,2017Natur.551...64A,2017ApJ...848L..27T,2017ApJ...850L...1L}, while X-ray and radio emission was detected 9 and 16 days post-merger respectively \citep{2017Natur.551...71T,2017Sci...358.1579H}. The delay in detecting the X-ray and radio emission stems from different physical origins -- the optical kilonova is produced by the decay of elements produced via r-process nucleosynthesis in the merger, while the X-ray and radio emission originates from the relativistic outflow launched by the merger interacting with the surrounding medium \citep{2012ApJ...746...48M}. Although none have been detected to-date, the slow moving material in the kilonova may eventually give rise to late-time radio emission \citep[e.g.][]{2018ApJ...867...95H,2019MNRAS.487.3914K}, while the synchrotron emission from the relativistic outflow was later detected at optical wavelengths \citep{2018NatAs...2..751L,2019ApJ...883L...1F,2019ApJ...870L..15L}.

Continued radio monitoring of GW170817 revealed a lightcurve that gradually rose over the following months \citep{2017ApJ...848L..21A,2018Natur.554..207M,2018ApJ...856L..18M}, before peaking approximately 150 days post-merger \citep{2018ApJ...858L..15D} and rapidly fading \citep{2018ApJ...863L..18A,2018MNRAS.478L..18T,2018ApJ...868L..11M}. This comprehensive monitoring campaign with radio, optical and X-ray telescopes has allowed tight constraints to be placed on the spectral and temporal evolution of the non-thermal afterglow \citep{2019ApJ...886L..17H,2020arXiv200602382M,2020MNRAS.498.5643T,2020RNAAS...4...68H}. In turn, these constraints have enabled physical properties of the merger, including the energetics, circum-merger density and inclination angle, to be inferred \citep[e.g.][]{2018ApJ...867...18N}. Further constraints on the outflow geometry were obtained through Very Long Baseline Interferometry (VLBI) observations \citep{2018Natur.561..355M,2019Sci...363..968G}, while the non-detection of linearly polarised radio emission constrains properties of the magnetic field of the outflow \citep{2018ApJ...861L..10C}.

There has been extensive research into the detectability of the radio afterglow of compact object mergers \citep[e.g.][]{2014arXiv1405.6219F,2016ApJ...831..190H,2017MNRAS.471.1652L,2019A&A...631A..39D,2019MNRAS.488.2405G,2020MNRAS.498.2384L}. The joint detection of GW170817 and GRB170817A confirmed the relationship between neutron star mergers and short GRBs \citep{2019MNRAS.483..840B,2019ApJ...880L..23W} that has been predicted for decades \citep{1989Natur.340..126E,1992ApJ...395L..83N}, and therefore studies into the detectability of short GRB radio afterglows are also of relevance \citep[e.g.][]{2013MNRAS.435.2543G,2014PASA...31...22G,2015ApJ...806..224M,2015RAA....15..237Z}. That being said, the intricacies of this relationship are not yet clear, as the GRB accompanying GW170817 was an outlier compared to the standard population and the gamma-ray emission likely did not originate from the core of the jet \citep{2019MNRAS.486.1563M,2019MNRAS.483.1247M,2019A&A...628A..18S}.

In this work we build upon previous studies by quantifying prospects for detecting afterglows with all major current and future GHz-frequency radio facilities using observing strategies tailored to the specifications of each facility. We have expanded our study beyond current gravitational wave detectors, and compare the maximum detectable distance to the range of detectors that will be built over the coming decades. In Section \ref{sec:afterglow_modelling} we describe the afterglow model we use in this work, and summarise the other forms of radio emission that may be produced by mergers. In Section \ref{sec:radio_telescopes} we summarise potential observing strategies for radio follow-up and the ability of existing and planned radio telescopes to carry out those observations. In Section \ref{sec:gw_detectors} we outline the detection and localisation capabilities of existing and planned gravitational wave detectors. Section \ref{sec:prospects} includes a discussion of the benefits and limitations of radio follow-up, and quantifies prospects for the detection of radio counterparts using the previously described afterglow model and observing strategies.

\section{Afterglow Physics}
\label{sec:afterglow_modelling}
\subsection{Emission from a power-law jet}
To assess the detectability of radio afterglows we use the same power-law jet model as \citet{2020MNRAS.494.2449D}. The kinetic energy is distributed according to
\begin{equation}
E(\theta) = \frac{E_{\rm iso}}{1+(\theta/\theta_{j,c})^{3.5}},
\end{equation}
where $\theta$ is the polar angle from the jet axis, $\theta_{j,c}$ and $E_{\rm iso}$ are the half opening angle and isotropic-equivalent energy of the core of the jet respectively. Here we assume $\theta_{j,c}=0.05\,{\rm rad}$, $E_{\rm iso}=10^{52}\,{\rm erg}$ based on the afterglow of GW170817 \citep{2018Natur.561..355M,2019Sci...363..968G,2020arXiv200602382M,2020MNRAS.498.5643T}. The power law profile of $E\propto \theta^{-3.5}$ is motivated by the result of a hydrodynamic simulation by \cite{2021MNRAS.500.3511G}.
We use the standard synchrotron afterglow model of \citet{1998ApJ...497L..17S} to calculate the radio flux produced by the radial expansion of a jet into a uniform medium. The initial Lorentz factor of the jet is given by
\begin{equation}
\Gamma(\theta) = 1+\frac{\Gamma_{c}}{1+(\theta/\theta_{j,c})^5},
\end{equation}
where $\Gamma_c$ is the Lorentz factor of the jet's core. Based on the population of short GRB afterglows \citep{2015ApJ...815..102F} and the afterglow of GW170817 \citep[e.g.][]{2020MNRAS.498.5643T,2019ApJ...886L..17H,2020ApJ...899..105L} we assume that the electron energy distribution follows a power law with index $p=2.16$. We also assume the fraction of shock energy distributed to the electrons and magnetic field of the jet to be $\epsilon_e=0.1$ and $\epsilon_B=0.01$ respectively, based on the canonical GRB microphysics parameters \citep[e.g.][and references therein]{2014ARA&A..52...43B}, which we note are higher than those typically found for GW170817 \citep[see Table 1 of][]{2020MNRAS.494.2449D}. Our results rely on these parameter assumptions, which are biased towards GW170817-like events and therefore may not represent the general population of compact binary coalescences.

We initially prepared a set of afterglow lightcurves $S_{\nu,0}$ as a function of observer time $t_0$ for observing angles from 10--90\,$\deg$ for fixed $E_{\rm iso}$, $n_0$, $\theta_{j,c}$ and microphysics parameters. We then calculate $S_\nu(t)$ for different values of $n$ by using the scaling relations
\begin{equation}
    S_\nu(t) = \left(\frac{n}{n_0}\right)^{(p+1)/4}S_{\nu,0},~~~~~~~~t = \left(\frac{n}{n_0}\right)^{-1/3}t_0
\end{equation}
\citep[see][]{2002ApJ...568..820G,2012ApJ...749...44V}.
Here we have assumed that the emission frequency lies below the synchrotron cooling frequency and above the characteristic frequency of the lowest energy electrons and self-absorption frequency (i.e. $\nu_m,\nu_{s}\ll \nu \ll \nu_c$). Population modelling \citep[e.g.][]{2019A&A...631A..39D} suggests that this is a valid assumption for the GHz-frequency range, which is the focus of this work.

While we only explicitly consider fixed $E_{\rm iso}$ and microphysics parameters in this work, for completeness we note that the lightcurves obey
\begin{equation}
    \label{eq:flux_scaling}
    S_\nu(t) \propto E_{\rm iso}(\epsilon_{B}n)^{(p+1)/4}\epsilon_{e}^{p-1},~~~~~t=\left(\frac{n_0E_{\rm iso}}{nE_{{\rm iso},0}}\right)^{1/3}t_0,
\end{equation}
which can be used to scale the results of this work to generalised energetics and microphysics parameters. We note that this model does not consider synchrotron self-absorption, and as such, the predicted flux density is overestimated for $n>1$\,cm$^{-3}$ and $\theta_{\rm obs} < 10\,\deg$. We therefore do not consider parameters outside of these limits in this work.

To scale the flux density from the nominal emission frequency of our model, $\nu_0$, to a general observing frequency, $\nu$, we assume the emission obeys a simple power law with spectral index $\alpha=(1-p)/2$ \citep{1998ApJ...497L..17S}. Since the model flux density, $S_0$, is calculated in the rest frame of the merger we also apply the standard K-correction \citep[see e.g.][]{2017A&A...602A...5N}, resulting in a frequency-corrected flux density of
\begin{equation}
    S = \frac{S_0}{(1+z)^\alpha} \left(\frac{\nu}{\nu_0}\right)^{\alpha}
\end{equation}
where we convert between redshift, $z$, and luminosity distance, $D_L$, using the cosmological parameters in \citet{2013ApJS..208...19H} implemented in {\sc astropy.cosmology.WMAP9}. We emphasise that this model is by no means comprehensive -- it is a single, generalised, model for one of many components of the radio emission that may be produced by compact binary mergers \citep[e.g.][]{2015MNRAS.450.1430H}.

\subsection{Other forms of emission}
At early times ($t\lesssim 10$\,days post-merger) the dominant source of radio emission may be the `reverse shock' \citep{1995ApJ...455L.143S,2016ApJ...825...48R,2019MNRAS.489.1820L}, which propagates from the outflow towards the site of the merger. This emission is dependent on similar parameters to the forward shock, including the inclination angle. Reverse-shock emission has been detected in long GRBs \citep[e.g.][]{1999ApJ...522L..97K,2000ApJ...545..807K,2000ApJ...542..819K,2013ApJ...776..119L,2014ApJ...781...37P,2017ApJ...848...69A,2019ApJ...884..121L,2020MNRAS.tmp.1856R} but may be difficult to detect in short GRBs due to the required follow-up latency \citep{2018Galax...6..103L}. However, observations of GRB 160821B suggest that a reverse shock component is responsible for part of the early time radio emission \citep{2019MNRAS.489.2104T,2019ApJ...883...48L}. We stress that even if reverse shock emission may be difficult to detect as a distinct component, its contribution to the overall emission may be significant enough that forward-shock jet models (including the one we use in this paper) alone may not accurately describe the early-time lightcurve evolution.

While most mergers likely launch a jet, the fate of the jet is dependent on factors including the energetics of the merger and the density of the surrounding environment. As the jet propagates into the surrounding ejecta, it may form a cocoon, which will produce a distinct signal as it expands and breaks out of the ejecta \citep{2014ApJ...784L..28N,2014ApJ...788L...8M,2017MNRAS.471.1652L,2017ApJ...834...28N,2018MNRAS.473..576G}. Beyond the initial rise and decline, this geometry may also produce a double-peak in the lightcurve \citep{2018arXiv180508338B}.

Radio emission may also be detectable from kilonovae, which produced the early-time optical and infrared emission associated with GW170817 \citep{2017Natur.551...67P,2017Natur.551...80K,2017Natur.551...75S}. This emission is produced by the ejecta associated with the kilonova which expands at sub-relativistic velocity and is therefore expected to be fainter, and peak at later times, than the non-thermal synchrotron emission. The timescale of this peak ranges from months to decades and the peak luminosity is similarly uncertain \citep[e.g.,][]{2011Natur.478...82N,2013MNRAS.430.2121P,2018ApJ...867...95H,2019MNRAS.487.3914K,2020MNRAS.495.4981M}. The kilonova ejecta are expected to expand quasi-isotropically, and therefore this form of radio emission may be detectable from future mergers even when the emission associated with the relativistic ejecta is not detected due to beaming.

\begin{table*}
    \caption{Capabilities of existing and planned radio facilities including observing frequency ($\nu$), bandwidth ($\Delta\nu$), field of view ($\Omega$), angular resolution ($\theta$). See Section \ref{sec:radio_telescopes} for details.}
    \label{tab:radio_facilities}
    \centering
    \begin{threeparttable}
    \begin{tabular}{lcS[table-format=1.2]S[table-format=1.1]S[table-format=2.2]S[table-format=2.3]cc}
    \hline\hline
    Facility & Band & $\nu$ & $\Delta\nu$ & $\Omega$ & $\theta$ & Dec. limit\\
     & & {(GHz)} & {(GHz)} & {(deg$^2$)} & {(arcsec)} & (deg)\\
    \hline
    ATCA\tnote{a} &  C/X & 8.0 & 8.0 & 0.01 & 2 & $<+30$\\
    GMRT & B3 & 0.4 & 0.2 & 1.4 & 8 & $>-50$\\
    ~ & B4 & 0.7 & 0.3 & 0.4 & 4 & $>-50$\\
    VLA\tnote{b} & L & 1.5 & 1.0 & 0.12 & 6 & $>-30$\\
    ~ & S & 3.0 & 1.5 & 0.06 & 2.7 & $>-30$\\
    ~ & C & 6.0 & 4.0 & 0.01 & 1.3 & $>-30$\\
    \rule{0pt}{3ex}Apertif & L & 1.4 & 0.3 & 6 & 15 & $>-20$\\
    ASKAP & Band 1 & 0.9 & 0.3 & 30 & 15 & $<+30$\\
    MeerKAT & L & 1.4 & 0.7 & 0.8 & 7 & $<+30$\\
    \rule{0pt}{3ex}DSA-2000 & -- & 1.35 & 1.3 & 10.6 & 3.5 & $>-30$\\
    SKA-1 & Band 2 & 1.43 & 0.4 & 0.8 & 0.6 & $<+30$\\
    ngVLA & Band 1 & 2.4 & 2.3 & 0.13 & 0.002 & $>-30$\\
    \rule{0pt}{3ex}SKA-2 & Band 2 & 1.43 & 0.4 & 0.8 & 0.6 & $<+30$\\
    \hline\hline
    \end{tabular}
    \begin{tablenotes}\footnotesize
    \item[a] Assumes a 6\,km array configuration
    \item[b] Assumes B configuration
    \end{tablenotes}
    \end{threeparttable}
\end{table*}

\section{Radio Telescope Capabilities}
\label{sec:radio_telescopes}
In this section we outline the follow-up capabilities of radio interferometers. We exclude low-frequency ($\leq 300$\,MHz) telescopes from this discussion as our model does not account for synchrotron self-absorption and is therefore not accurate at these frequencies. Table \ref{tab:radio_facilities} lists the specifications of each telescope in our analysis.

\begin{figure}
    \centering
    \includegraphics{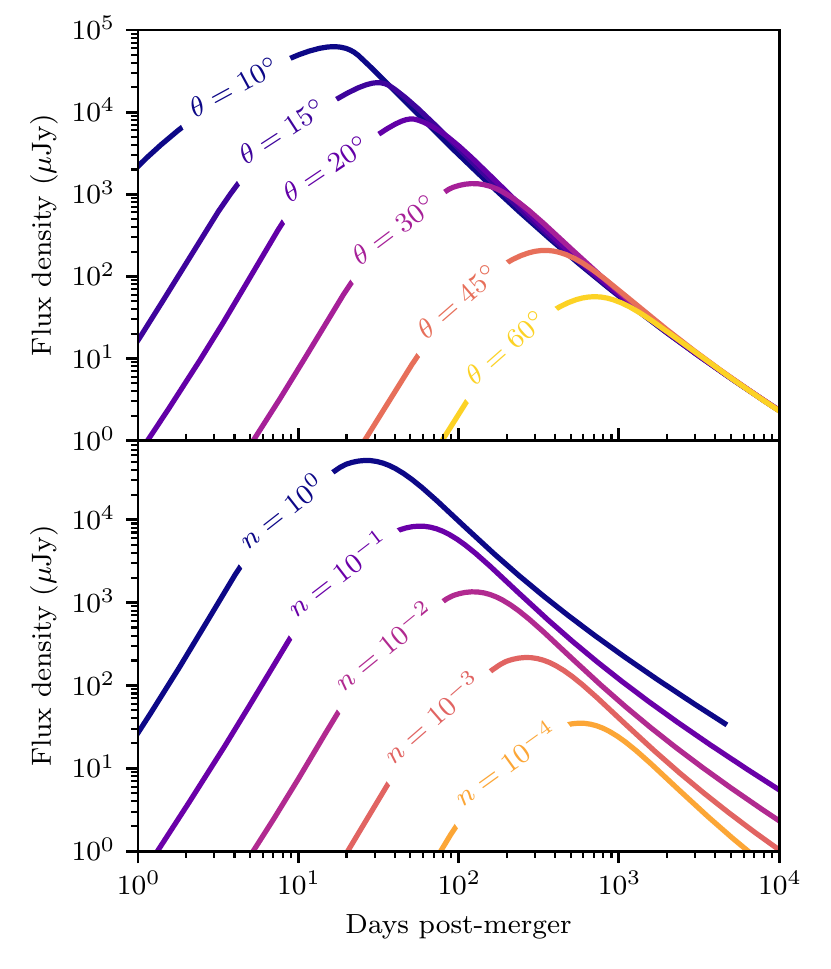}
    \caption{Comparison of lightcurve properties for a range of inclination angle, $\theta$, and circum-merger density, $n$ (in units of cm$^{-3}$), assuming a merger distance of 40\,Mpc and an observing frequency of 1.4\,GHz for the merger parameters outlined in Section \ref{sec:afterglow_modelling}. Top: Lightcurves for a range of inclination angles, with $n=10^{-2}\,$cm$^{-3}$, corresponding to typical short GRB circum-burst density. Bottom: Lightcurves for a range of circum-merger densities, with $\theta=30\,\deg$.}
    \label{fig:lightcurve_comparison}
\end{figure}

\subsection{Follow-up Strategies}
\label{subsec:strats}
The localisation areas of previously detected gravitational wave events span tens to thousands of square degrees \citep{2019PhRvX...9c1040A,2020arXiv201014527A} and while the addition of more detectors to the network will lead to smaller localisation areas for events at comparable distances, improvements to detector sensitivity will also lead to the detection of more distant events with comparable localisation areas \citep{2020LRR....23....3A,2019CQGra..36v5002H}. Therefore the largest impediment to detecting an electromagnetic counterpart is the ability to cover sufficient area to a sufficient depth. Below we outline possible observing strategies for detecting a long-lasting (months-years), sub-mJy radio afterglow as predicted by theoretical models \citep[e.g.][]{2016ApJ...831..190H} and observed from GW170817 \citep[e.g.][]{2019ApJ...886L..17H,2020arXiv200602382M,2020MNRAS.498.5643T}. The search for other forms of emission (e.g. an optical counterpart or a coherent radio burst) may require different strategies. 

\subsubsection{Targeting known galaxies}
\label{subsubsec:galaxy_targeting}
One survey strategy adopted in the follow-up of GW170817 was to target known galaxies in the localisation volume \citep[e.g.,][]{2017PASA...34...69A,2017Natur.551...64A,2017Sci...358.1556C,2017ApJ...848L..29D,2017Sci...358.1565E,2017ApJ...848L..24V}. This was also the predominant strategy used in radio follow-up, as at the time no facilities had sufficient survey speed to perform an unbiased search for a radio counterpart.

Currently the best catalogues for this purpose include the Galaxy List for the Advanced Detector Era \citep[GLADE;][]{2018MNRAS.479.2374D}, Census of the Local Universe \citep[CLU;][]{2019ApJ...880....7C}, Photometric Redshifts for the Legacy Surveys \citep[PRLS;][]{2021MNRAS.501.3309Z}, the WISE $\times$ SuperCOSMOS Photometric Redshift Catalog \citep{2016ApJS..225....5B}, and the 2MASS photometric redshift galaxy catalogue \citep[2MPZ;][]{2014ApJS..210....9B}. GLADE and CLU are compilations of existing surveys (and encompass observations from 2MPZ), while PRLS uses the Dark Energy Camera Legacy Survey \citep{2019AJ....157..168D} and data from the \textit{Wide-field Infrared Survey Explorer} \citep[WISE;][]{2010AJ....140.1868W}. Additionally, the Mass AssociatioN for GRavitational waves ObserVations Efficiency \citep[MANGROVE;][]{2020MNRAS.492.4768D} combines GLADE with estimates of stellar mass from WISE, and \citet{2020MNRAS.495.1841A} compute host probabilities from estimates of stellar mass and star formation rate.

Numerous works explore the prospects for this search method -- for example, \citet{2016ApJ...820..136G} demonstrates that simple galaxy targeting can reduce the required number of pointings by a factor of 10--100 compared to an unbiased search, while \citet{2018MNRAS.478..692C} find that search efficiency can further be improved by a factor of 2. Some strategies expand on basic galaxy targeting by preferring to target galaxies that are closer \citep{2017ApJ...848L..33A}, or by using assumptions on afterglow emission \citep{2016ExA....42..165C,2017ApJ...846...62S}. Telescopes with wider fields of view may be able to observe multiple galaxies in a single pointing, and this follow-up may be optimised by convolving the gravitational wave skymap with a galaxy catalogue \citep{2016MNRAS.462.1591E}.

Once we have obtained a larger sample of localised mergers we will be able to use the population of known host galaxies to weight galaxy catalogues by more robust metrics than simple mass-analogues. While observational constraints are currently limited by a lack of events, detailed studies of the host galaxy of GW170817 were carried out \citep{2017ApJ...848L..22B,2017ApJ...849L..16I,2017ApJ...848L..28L,2017ApJ...848L..30P,2020A&A...634A..73E,2018A&A...620A..37C} and simulations are already revealing potential relationships between merger and host properties \citep[e.g.][]{2019MNRAS.489.4622T,2019MNRAS.487.1675A,2020MNRAS.491.3419A}. Radio observations will have a role to play in these efforts, as they are useful in determining the presence of AGN and thereby determine whether stellar mass or star formation is the driving factor of merger formation \citep{2007ApJ...665.1220Z,2017arXiv171106873C}. They can also measure star formation rate in galaxies with optical obscuration \citep{2013ApJ...778..172P}.

Ultimately, the limiting factor of this technique is the completeness of the survey catalogue, which for current surveys \citep[e.g.][]{2018MNRAS.479.2374D,2019ApJ...880....7C} is lacking at distances comparable to the planned LIGO detector range of $330\,$Mpc \citep{2020LRR....23....3A}. While PRLS is deeper than the other wide surveys (with a limiting magnitude of $\sim 20.4$ subject to other selection criteria), it only covers $\sim 16\%$ of the sky and therefore cannot be utilised for all events. We discuss the issue of galaxy catalogue completeness and its impact on follow-up in \ref{subsec:galaxy_cat_completeness}. For more distant events it is therefore natural to consider searches that only use the localisation obtained from the gravitational wave signal.

\subsubsection{Unbiased searches}
Telescopes with larger survey speeds are capable of performing unbiased searches for radio counterparts by observing entire localisation regions. Here we consider unbiased searches to be feasible for telescopes with a survey speed larger than $10\,\deg^2\,$hr$^{-1}$ for a detection threshold of 0.5\,mJy. This choice reflects the requirement to cover a substantial fraction of a localisation area (typically tens--hundreds of square degrees) to a constraining sensitivity (given that afterglows are expected to peak at tens--hundreds of $\mu$Jy) in a reasonable amount of time. This technique has the obvious advantage of not being limited by the completeness of existing galaxy catalogues, while on the other hand, can result in coverage of a large amount of extraneous area and an increase in false-positives (which we discuss in Section \ref{subsec:false_positives}). However, this can be partly mitigated by restricting candidates to those associated with known galaxies within the localisation volume.

There are multiple ways to optimise an unbiased search \citep[e.g.][]{2016A&A...592A..82G,2017PASP..129k4503G,2019PASA...36...19D,2020PhRvD.101l3008G} but for the purposes of this work we simply assume that a telescope can observe a given sky area without any consideration of exact observing strategies.

\subsubsection{Monitoring known electromagnetic counterparts}
The final type of follow-up involves searching for radio emission from known counterparts detected at other wavelengths, which allows for maximised sensitivity as all available telescope time can be allocated to a single pointing.

The detection of a known counterpart can also help inform the required cadence and sensitivity of radio observations. Localising the counterpart to a galaxy provides accurate distance measurements \citep{2017ApJ...848L..31H} which can also be used to infer tighter constraints on the merger inclination angle independent of any afterglow modelling \citep{2018ApJ...853L..12M}. The detection of a short GRB can loosely constrain the merger energetics \citep{2017ApJ...848L..13A,2017ApJ...848L..14G}, and a combination of multi-wavelength detections can be used to broadly infer the overall geometry of the merger \citep{2017Sci...358.1559K}.  

Even if a radio counterpart is not detected, radio observations are still useful in placing constraints on the spectral properties of the synchrotron afterglow (in combination with optical and X-ray follow-up) and inferring properties of the host galaxy. Single-dish spectral line observations can also be used to constrain the average density of the host galaxy and infer the density of the environment surrounding the merger \citep{2017Sci...358.1579H}. X-ray observations can also be used as independent confirmation of these estimates \citep[e.g.][]{2019ApJ...886L..17H}.

\subsection{Existing Facilities}
\label{subsec:existing_facilities}
\subsubsection{Australia Telescope Compact Array (ATCA)}
The Australia Telescope Compact Array is an East-West array consisting of six 22-m dishes with a maximum baseline of 6\,km (depending on the array configuration). Each dish is equipped with a set of receivers that can sample frequencies from 1.1 to 105\,GHz, with the System Equivalent Flux Density (SEFD) increasing towards higher frequencies \citep{2011MNRAS.416..832W}. In this paper we consider observations in the C/X band (4--12\,GHz) which is generally the most sensitive due to the high Radio Frequency Interference (RFI) occupancy at lower frequencies\footnote{\url{https://www.narrabri.atnf.csiro.au/observing/rfi/monitor/rfi_monitor.html##atca}}.

Currently correlation is performed using the Compact Array Broadband Backend \citep{2011MNRAS.416..832W}, which allows observations in $2\times2048$\,MHz windows. However, the Broadband Integrated-GPU Correlator for the Australia Telescope (BIGCAT) upgrade, which will be completed by December 2021, will double the available bandwidth to 8\,GHz. As this upgrade will occur on a similar timeline to the start of the next LIGO/Virgo observing run, in this paper we assume that the upgrade is successfully commissioned and therefore the sensitivity will improve by a factor of $\sqrt{2}$ compared to the current level.

It is not feasible to carry out unbiased follow-up with the ATCA due to its low survey speed, even using mosaicing. However, it is useful for targeted follow-up, either of galaxies within the localisation volume of the gravitational wave event, or of counterparts detected at other wavelengths. Taking into account observing overheads, it's possible to observe $\sim 50$ galaxies to a detection threshold of $<70\,\mu$Jy in a typical 12-hour observation\footnote{The ATCA is an East-West array and requires a full 12 hour track for complete uv coverage.} (12 minutes per source), while a detection threshold of $\sim 15\,\mu$Jy can be achieved for follow-up of a single source. We note that in many cases this can be improved by optimising pointings such that multiple galaxies are observed within the same field of view\footnote{\url{https://github.com/ddobie/atca-ligo}}, but to simplify the estimates in this paper we assume that we can simply observe 50 galaxies with 1 per pointing.

\subsubsection{Karl. G. Jansky Very Large Array (VLA)}
The Karl. G. Jansky Very Large Array (VLA) consists of twenty seven 25-m antennas with maximum baselines ranging from 36.4\,km (A configuration) to 1.03\,km (D configuration). In this paper we consider the B configuration which has baselines spanning 0.21--11.1\,km, although the A and C (which have maximum baselines of 36.4 and 3.4\,km respectively) configurations would also be suitable . The telescope can observe at frequencies from 58\,MHz to 50\,GHz, but in this case we consider observations at S-band (3\,GHz; offering the best sensitivity for the expected negative spectral index) and L-band (1.5\,GHz; offering lower sensitivity but a twice-larger field of view).

\citet{2019arXiv190407335R} propose an optimised galaxy targeting strategy for the VLA, improving the probability of detecting a radio counterpart by a factor of two compared to a simple approach. This strategy will enable observers to cover approximately 200 galaxies to a sensitivity of $15\,\mu$Jy. The VLA can also carry out unbiased searches using on-the-fly mosaicing, where the antennas are driven at a constant rate along a strip of sky, eliminating slew overheads which are dominated by the settling time of $\sim 7$ seconds per pointing. This technique has already been applied to follow-up of two gravitational wave events, GW151226 \citep{2016PhRvL.116x1103A} and GW190814 \citep{2020ApJ...896L..44A} covering $100\,\deg^2$/10\% \citep{2018ApJ...857..143M} and $5\,\deg^2$/50\% \citep{2019GCN.25690....1M} of the localisation regions respectively.

\subsubsection{Giant Metrewave Radio Telescope (GMRT)}
The Giant Metrewave Radio Telescope (GMRT) consists of thirty 45-m antennas with a maximum baseline of 25\,km. Recent upgrades to the receivers give the GMRT unparalleled sensitivity at sub-GHz frequencies, and here we consider observations in bands 3 and 4, centered on 400 and 700\,MHz respectively. Observations in these lower bands are vital in understanding the spectral evolution of the afterglow, in particular, placing constraints on the evolution of the synchrotron self-absorption frequency.

Here we consider the GMRT as a dedicated follow-up instrument for localised mergers, as its combination of sensitivity and field of view is not conducive to unbiased searches. We consider 3 hour observations in both Band 3 and 4, corresponding to an image sensitivity of 15 and 20$\,\mu$Jy respectively. This choice of exposure time is to ensure both bands can be observed in a typical GMRT session of $\sim 6$ hours.

\subsubsection{Arcminute Microkelvin Imager Large Array (AMI-LA)}
The Arcminute Microkelvin Imager Large Array (AMI-LA) consists of eight 12.8m antennas with a maximum baseline of 110\,m operating at an observing frequency of 15\,GHz. The AMI-LA Rapid Response Mode \citep[ALARRM;][]{2013MNRAS.428.3114S} enables the telescope to respond to GRB alerts within 2 minutes of the burst, allowing for tight constraints to be placed on early-time emission \citep[e.g.][]{2019MNRAS.489.1820L}. However, the AMI-LA field of view limits the practicality of early-time follow-up of neutron star mergers to those with a simultaneous detection of a GRB by the \textit{Neil Gehrels Swift Observatory}. While AMI-LA can also be used to monitor known counterparts, providing additional spectral coverage and achieving a sensitivity of $25\,\mu$Jy in a 3 hour observation, we do not consider it in our detectability analysis in Section \ref{subsec:follow_up_detectability}, and omit it from Table \ref{tab:radio_facilities}, due to the superior sensitivity of the VLA at these frequencies.

\subsubsection{Australian Square Kilometre Array Pathfinder (ASKAP)}
The Australian Square Kilometre Array Pathfinder \citep[ASKAP;][]{2008ExA....22..151J,2021PASA...38....9H} is an array of thirty-six 12-m dishes with baselines ranging from 37\,m to 6\,km. ASKAP is designed for all-sky surveys between 700-1800\,MHz (in this case we consider observations at 900\,MHz to maximise the sensitivity) and uses MkII phased-array feeds \citep{HampsonPAF} consisting of 36 beams resulting in a $30\,\deg^2$ field of view. While the SEFD is higher than other comparable telescopes, the large field of view results in a high survey speed, making it possible to search large areas of sky for a gravitational wave counterpart. However, this does mean that ASKAP is not useful for monitoring events where an electromagnetic counterpart has already been discovered. We therefore only consider the utility of ASKAP for localising events, and not monitoring them.

Similarly, ASKAP is not particularly useful for a galaxy-targeted search strategy since tens--hundreds of candidate host galaxies can be covered with a single pointing. Some widefield telescopes select their pointing strategy by convolving the localisation skymap with galaxy catalogues \citep[e.g.,][]{2016MNRAS.462.1591E} but in \citet{2019PASA...36...19D} we demonstrated that this strategy does not produce any appreciable benefits for ASKAP follow-up. In this work we consider two strategies, a single deep $\sim 10$ hour pointing (achieving a noise of $35\,\mu$Jy) or a widefield strategy consisting of four $\sim 3$ hour pointings (achieving a noise of $110\,\mu$Jy) for events localised to $<100\,\deg^2$ \citep[corresponding to $\sim 60\%$ of neutron star mergers during O4, and the majority of mergers detected with 3G detectors][]{2020LRR....23....3A,2019CQGra..36v5002H}. So far ASKAP has performed follow-up of S190510g and GW190814 \citep[][Stewart et al. in prep.]{2019ApJ...887L..13D} and both times the latter pointing strategy was used. We note that due to the comparably low angular resolution of ASKAP, some afterglows may be contaminated by emission from the host galaxy. We discuss this problem in Section \ref{subsec:radio_loud_hosts}.

\subsubsection{Apertif}
The Westerbork Synthesis Radio Telescope (WSRT) is an East-West array consisting of fourteen 25-m antennas with a maximum baseline of 2.7\,km. Twelve of the antennas have been fitted with L-band (operating from 1--1.75\,GHz) PAFs as part of the \textit{APERture Tile In Focus} (Apertif) project, improving the telescopes field of view to $9.5\,\deg^2$ \citep{2010iska.meetE..43O,2019NatAs...3..188A}. Similar to ASKAP, Apertif's wide field of view means it is more suited to widefield searches rather than targeted follow-up, while the observing frequency and maximum baseline mean it is also subject to the same host galaxy contamination issues. The sensitivity of Apertif is comparable to ASKAP, and we consider follow-up with one deep 12 hour pointing achieving a sensitivity of 25\,$\mu$Jy or four pointings achieving a sensitivity of 50\,$\mu$Jy.

\subsubsection{MeerKAT}
MeerKAT \citep{2016mks..confE...1J} consists of sixty four 13.5-m diameter antennas with a maximum baseline of 8\,km fitted with 1.4\,GHz receivers and has a $\sim 1\,\deg^2$ field of view. The highly sensitive receivers have a system equivalent flux density of $\sim 430\,$Jy \citep{2020ApJ...888...61M}, making MeerKAT a suitable facility for searching for emission from known counterparts. Here we consider two follow-up strategies -- a single deep 10\,hour pointing achieving a sensitivity of $\sim 2\,\mu$Jy and ten 1\,hour pointings covering $10\,\deg^2$ to a sensitivity of $\sim 7\,\mu$Jy. The former strategy applies to very well localised events (comparable to expectations for third generation detectors, see \ref{subsec:3g_detectors}) and monitoring of known counterparts, while the latter strategy applies to unbiased follow-up of events localised to $\lesssim 10\,\deg^2$ or targeted follow-up of less-localised events using a galaxy catalogue convolution strategy \citep{2016MNRAS.462.1591E}.

\subsection{Future Facilities}
\label{subsec:future_facilities}
\subsubsection{Square Kilometre Array (SKA)}
The Square Kilometre Array (SKA) will be the worlds largest radio telescope, with a mid-frequency array in South Africa and a low-frequency array in Western Australia. In this paper we focus on the mid-frequency array, which will be split into two stages.

SKA-1 mid will consist of the existing MeerKAT array and an additional one hundred and thirty three 15-m dishes, with the array expected to come online in the mid-2020s. \citet{2019arXiv191212699B} outlines the anticipated array performance, and for the purposes of this paper we consider observations in Band 2, centered on 1.43\,GHz as a compromise between maximising sensitivity and the expected negative spectral index of gravitational wave afterglows. We assume the same observing strategy as MeerKAT, with a 1 hour continuum sensitivity of $2\,\mu$Jy.

We also consider observations with the SKA-2. While the design specifications for the SKA-2 are still uncertain, here we assume an order of magnitude sensitivity improvement over SKA-1 for the same observing strategies.

\subsubsection{Next Generation Very Large Array (ngVLA)}
The Next Generation Very Large Array \citep[ngVLA;][]{2018ASPC..517....3M} is a planned replacement for the VLA operating between 1.2 and 116\,GHz. The main array will consist of two hundred and fourteen 18-m dishes on baselines of up to 1000\,km. We consider observations in the lowest frequency band, centered on 2.4\,GHz due to the larger field of view and higher relative sensitivity \citep{2018ASPC..517...15S}. It is currently anticipated that Early Science observations will begin in 2028 with full operation from 2034 onwards\footnote{\url{https://ngvla.nrao.edu/page/faq\#faq_16_content}}.

The ngVLA will be capable of performing unbiased follow-up of well-localised events, observing $10\,\deg^2$ to a sensitivity of $1\,\mu$Jy in 10 hours \citep{2019arXiv190310589C}. The ngVLA has comparable instantaneous sensitivity to the SKA, making it the premier northern hemisphere facility for targeted follow-up of known counterparts. The ngVLA is also capable of observing at frequencies up to 93\,GHz, compared to the SKA which will only observe up to 12.5\,GHz \citep{2019arXiv191212699B}. This will allow a better characterisation of the spectral properties of afterglows in conjunction with optical and X-ray telescopes, as was performed for GW170817 \citep{2020arXiv200602382M,2020MNRAS.498.5643T}.

\subsubsection{Deep Synoptic Array-2000 (DSA-2000)}
The Deep Synoptic Array (DSA-2000) is a proposed telescope that will consist of two thousand 5-m dishes capable of simultaneously observing between 0.7-2\,GHz and is scheduled to be fully operational by 2026 \citep{2019BAAS...51g.255H}. The telescope is optimised for survey speed, with a 10.6\,deg$^2$ instantaneous field of view and 2.5\,Jy SEFD. Approximately 1 hour per day will be allocated to follow-up of gravitational wave events, with the focus being on well-localised events that can be covered with a single pointing. In these cases, the entire localisation region can be covered to a $5\sigma$ detection threshold of $\sim 5\,\mu$Jy. Similar to ASKAP, we also consider a widefield strategy of twelve $\sim 5$ minute pointings for follow-up of events localised to $\sim 100\,\deg^2$.

While the DSA-2000 can cover a $10\,\deg^2$ localisation ten times faster than the SKA or ngVLA, it is important to note that both telescopes can cover that area more efficiently. Gravitational wave localisations are generally irregular shapes (although this will be less true as we move to arrays with >3 detectors) and often multi-modal, while telescope fields of view are either circular or rectangular. \citet{2016A&A...592A..82G} find that it is significantly more efficient to cover localisation regions with a distributed group of multiple small-FoV telescopes than a single widefield telescope due to the lower extraneous coverage. Similarly, the SKA/ngVLA will use many small-FoV pointings and will therefore achieve more efficient coverage than the DSA-2000 strategy of fewer large-FoV pointings.

\subsection{Serendipitous Observations}
\label{subsec:serendipitous}
Advances in radio telescope technology will allow for numerous widefield surveys of the radio sky to be undertaken in the coming decades. Deep all-sky surveys will provide sensitive reference images for transient follow-up, and widefield transient searches will likely provide serendipitous coverage of gravitational wave events.

\subsubsection{ASKAP}
The Evolutionary Map of the Universe \citep[EMU;][]{2011PASA...28..215N} will cover the sky South of $+30\,\deg$ declination to a design sensitivity ($1\sigma$ noise) of $\sim 10\,\mu$Jy at 1.3\,GHz. This will provide the most sensitive map of the Southern radio sky to-date and be useful as a reference image for transient searches. The Rapid ASKAP Continuum Survey \citep[RACS;][]{2020PASA...37...48M} achieves a typical sensitivity of $250\,\mu$Jy at $900\,$MHz and has already been used as a reference image in follow-up of GW190814 \citep{2019ApJ...887L..13D}.

The proposed Variables And Slow Transients \citep[VAST;][]{2013PASA...30....6M} survey is split into three main components, Wide, Deep and Galactic, and will span at least 5 years. VAST-Wide will observe an area of $10000\,\deg^2$ to a detection threshold of $2.5\,$mJy on a daily cadence. VAST-Deep will achieve a detection threshold of $250\,\mu$Jy and will observe $10000\,\deg^2$ 7 times, and a single $30\,\deg^2$ field daily. VAST-Galactic will observe $750\,\deg^2$ of the galactic plane 64 times to a detection threshold of $500\,\mu$Jy, which will be useful for events where optical follow-up is hindered by extinction and a high rate of unrelated transients. Here we consider the VAST-Wide and low cadence VAST-Deep surveys, as the wide areal coverage makes them more conducive to this kind of search. We also note that the ASKAP Survey Science observing strategy has not yet been finalised

\subsubsection{MeerKAT}
While MeerKAT has a higher instantaneous sensitivity than ASKAP, there are currently no plans to use it for a dedicated widefield untargeted transients survey \citep{2017arXiv171104132F}. Instead, transient searches will be conducted using commensal data from other surveys, which cover relatively small areas of sky \citep[generally less than 30$\deg^2$, see][]{2012IAUS..284..496H,2016mks..confE...8S,2016mks..confE...7D,2018arXiv180307424B}. Therefore the proposed transients search with MeerKAT will search a smaller area of sky to a greater sensitivity compared to ASKAP. As this is less conducive to serendipitous coverage of multi-messenger events, we instead consider an idealised untargeted survey for radio transients covering 5000\,$\deg^2$ observed to a sensitivity of $20\,\mu$Jy with 9 observations separated by 4 months. This corresponds to a total observing time of 3750 hours.

\subsubsection{Deep Synoptic Array}
The Cadenced All-Sky Survey \citep{2019BAAS...51g.255H} will observe 16 epochs of the sky North of $-30\,\deg$ on a 4-month cadence to a detection threshold of $10\,\mu$Jy, providing coverage of the majority of gravitational wave events. It will also ultimately provide a reference image with an rms noise of 500\,nJy.

\begin{table}
    \caption{Capabilities of gravitational wave detector networks made of the Hanford (H), Livingston (L), Virgo (V), Kagra (K), LIGO-India (I) detectors. Detectors improved by the A+ upgrade are denoted by a subscript $+$ while LIGO-Voyager detectors are denoted by a subscript $V$.}
    \label{tab:gw_facilities}
    \centering
    \begin{threeparttable}
    \begin{tabular}{llcccc}
    \hline\hline
    Epoch & Facilities & Timeline & Range\tnote{a} & Localisation\tnote{b} & Rate\tnote{c}\\
     & & & (Mpc) & ($\deg^2$) & (yr$^{-1}$)\\
    \hline
    O4 & HLVK & 2022--23 & 190\tnote{d} & 35 & 10\\
    O5 & H$_{+}$L$_{+}$V$_{+}$K & 2025--26 & 330\tnote{d} & 35 & 50\\
    2G & H$_{+}$L$_{+}$V$_{+}$KI$_{+}$ & 2026 & 330 & 35 & 50\\
    \rule{0pt}{3ex}Voy. & H$_{\rm V}$L$_{\rm V}$V$_{\rm V}$ & 2030 & 1100 & 70 & 1800\\
    \rule{0pt}{3ex}3G & ET, CE, Voy & 2040 & $5\times10^{4}$ & 10 & $10^8$\\
        ~ & ET, 2CE & & $5\times10^{4}$ & 1 & $10^8$\\
    \hline\hline
    \end{tabular}
    \begin{tablenotes}\footnotesize
    \item[a] Maximum range of any detector in the network
    \item[b] Order of magnitude estimate for typical localisation
    \item[c] Number of detections per year assuming a merger rate of $320\,$Gpc$^{-3}$yr$^{-1}$ \citep{2020arXiv201014533T}
    \item[d] We assume a total network range of 160\,Mpc and 300\,Mpc for O4 and O5 respectively
    \end{tablenotes}
    \end{threeparttable}
\end{table}

\section{Gravitational Wave Detectors}
\label{sec:gw_detectors}
\subsection{Second Generation Detectors}
\subsubsection{Fourth Observing Run (O4; 2022--2023)}
The Fourth Observing Run (O4) will run for one year with both LIGO detectors close to the design sensitivity of a sky and inclination angle averaged detection range of 190\,Mpc \citep{2020LRR....23....3A}. Advanced Virgo will have a binary neutron star range of 90--120\,Mpc (comparable to the LIGO detector ranges during O3), while the sensitivity of KAGRA \citep{2019NatAs...3...35K,2020arXiv200802921K} has a large uncertainty and estimates for its binary neutron star range is 25--130\,Mpc. Assuming a KAGRA range of 80\,Mpc the estimated number of detections is $10^{+52}_{-10}$, with a median 90\% localisation of $33\,\deg^2$. We adopt a detection range of 160\,Mpc based on the minimum specifications for the LIGO detectors and reflecting the lower range of the other two detectors.

\subsubsection{Fifth Observing Run (O5; 2024--2025)}
The fifth observing run will begin after the A+ upgrade, which will increase the LIGO detector range to 330\,Mpc \citet{2020LRR....23....3A}. The Virgo detector will also undergo significant upgrades, and will operate with a binary neutron star range of 150--260\,Mpc and KAGRA will operate with a range of at least 130\,Mpc and possibly as high as 155\,Mpc. For the purposes of this paper we assume a sky and inclination angle averaged detection range of 300\,Mpc.

\subsubsection{A Five Detector Network (2025+)}
LIGO-India \citep{M1100296-v2} is expected to join operations in 2025 and will eventually reach design specifications with a range of 330\,Mpc. The geographical location of LIGO-India improves the localisation capabilities of the global network by a factor of 5--10, and in some cases may result in mergers being localised to areas as small as $1\,\deg^2$.

\subsection{Third Generation Detectors}
\label{subsec:3g_detectors}
LIGO Voyager \citep{T1900409-v5} is a planned upgrade to the three existing LIGO facilities that will increase their range to 1.1\,Gpc, and also improve the localisation of closer events. This upgrade is expected to occur by the end of the decade, and we assume an operational start date of 2030. Voyager is an intermediate step between the Advanced LIGO detectors and third generation detectors, however due to the significant improvements in the nominal detector range and the uncertainty in the design and timeline of the detector, we consider it a third generation detector for the purposes of this work.

The proposed Neutron Star Extreme Matter Observatory \citep[NEMO;][]{2020PASA...37...47A} will also bridge the gap between detectors like A+ and true third-generation detectors. While the addition of this detector to the network will not significantly improve detector horizons or merger localisations, it will make the detection of post-merger gravitational waves feasible by extending the observable frequency range. This will provide insight into the nature of merger remnants \citep{2019ApJ...875..160A}, and the properties of the jet produced by the merger \citep[e.g.,][]{2020ApJ...895L..33B}. Constraints on both of these will inform radio follow-up efforts.

True third generation gravitational wave detectors will be an order of magnitude more sensitive than current detectors due to reduced quantum shot noise, improved mirror coatings and the placement of the detectors deep underground to reduce Newtonian noise. Both proposed detectors will be triple Michelson interferometers \citep{2009CQGra..26h5012F}, as opposed to the L-shaped interferometers used in second generation detectors, which enables the measurement of the gravitational wave polarisation.

The Einstein Telescope \citep[ET;][]{2010CQGra..27s4002P} will have 10\,km arms, while Cosmic Explorer (CE) will have 40\,km arms resulting in a higher sensitivity \citep{2017CQGra..34d4001A}. While a single third generation detector will be capable of detecting compact binary coalescences, precise localisation requires multiple detectors. \citet{2019CQGra..36v5002H} outline various observing scenarios based on combinations of Voyager, ET and CE detectors. For the purposes of this work we consider three simplified scenarios - the proposed network of three Voyager detectors, one Voyager detector with two 3G detectors (likely both ET and CE) and three 3G detectors (likely ET and two CEs). 
A network of one ET detector and two CE detectors will localise $>10\%$ of neutron star coalescences at redshift $z=0.3$ (corresponding to $\gtrsim 50$ events per year) to $\sim 0.1\,\deg^2$ with a median localisation of $1\,\deg^2$ \citep{2019CQGra..36v5002H}.

\section{Searching for Radio Afterglows}
\label{sec:prospects}
\subsection{Benefits and Limitations}
\subsubsection{Advantages over other wavelengths}
The ultraviolet/optical/infrared luminosity of kilonovae is dependent on parameters including the mass and velocity of the ejecta and the fraction of lanthanides produced \citep[e.g. see][and references therein]{2019LRR....23....1M}. Additionally, in higher mass mergers (including neutron star-black hole mergers), the blue component of the kilonova that was vital in localising GW170817 may not be produced and even the red component may be suppressed \citep{2013ApJ...778L..16H,2017Natur.551...80K}. The lack of optical counterparts detected during O3 suggests that not all events will produce kilonovae comparable to GW170817 \citep{2020ApJ...905..145K}. Radio afterglows are mostly independent of these parameters, and therefore probe an independent part of the merger parameter space to optical observations.

While the flux density of a radio afterglow is dependent on viewing angle, even significantly off-axis mergers will produce detectable emission (see Figure \ref{fig:lightcurve_comparison}). In comparison, short gamma-ray bursts are highly anisotropic and therefore most mergers will not produce a detectable GRB counterpart. In fact, the true rate of GRBs may be as much as $10^4$ times higher than what we detect \citep{2006ApJ...638..930S}, and radio observations present a promising method of detecting afterglows from off-axis events \citep{2002ApJ...576..923L}. However, the inclination angle dependence of GW detector sensitivity means that the fraction of GW events with a GRB counterpart is higher than the fraction of detectable GRBs from the general population, and the rate of joint GW-GRB detections will be a few per year for O4 and tens per year for designed LIGO specifications \citep{2019MNRAS.485.1435H,2020MNRAS.493.1633S}.

Extrinsic factors such as dust extinction and solar angle may also limit follow-up at other wavelengths. For example, comprehensive optical follow-up of GW170817 would not have been possible had the merger occured a month later, and even X-ray monitoring was hindered by $\sim 90$ days of sun avoidance. Radio telescopes are not limited by either of these factors in general, although some may have worse performance pointing near the Sun.

\subsubsection{False-positive rate}
\label{subsec:false_positives}
Radio searches are limited by the discovery of false positives in the form of variable sources manifesting as transients, as well as unrelated transients. Previous untargeted searches for radio transients over long ($>1\,$d) timescales \citep[][O'Brien et al. in prep., Stewart et al. in prep.]{2013ApJ...768..165M,2016ApJ...818..105M} have been dominated by variable Active Galactic Nuclei (AGN). Some variable AGN will be straightforward to immediately classify, but there exist many that cannot be classified without comprehensive broadband observations \citep[e.g.][]{2006MNRAS.371..898S,2008ApJ...689..108L,2021IAUS..359...27N}. Searches will also be hindered by sources exhibiting extrinsic variability caused by interstellar scintillation, which can cause variabiliy of tens of percent at GHz frequencies \citep{2002astro.ph..7156C}, resulting in the same compact source being undetected in one epoch and detected in the next. This can be mitigated by comparing candidate counterparts to galaxy catalogues and ruling out any that are spatially consistent with the nucleus --  neutron star mergers are likely to occur away from the galaxy nucleus. However, some telescopes may not provide sufficient astrometric accuracy to do this, necessitating follow-up observations with other facilities.

Searches consisting of multiple short-integration pointings may also discover short-duration radio transients such as Fast Radio Bursts, or flare stars \citep[][]{1999AJ....117.1568H,2019ApJ...871..214V,2021MNRAS.502.5438P}, although these will be easily ruled out as unrelated by comparison to archival data along with subsequent follow-up observations.

As radio follow-up observations become more sensitive, we will also discover afterglows from other transients including tidal disruption events and a variety of GRBs. At a detection threshold of $10\,\mu$Jy the expected areal density of radio transients is a few per $\deg^2$ at GHz frequencies \citep{2015ApJ...806..224M}. Since the emission from most radio transients originates from a synchrotron blast wave, even broadband radio observations may not be sufficient to classify transient types, and instead long-term monitoring to determine the temporal evolution of the source, along with multi-wavelength follow-up, will be required.

While widefield optical searches discover thousands of false positives, ruling them out is made easier by having a large sample of known optical transients with more distinct spectra and underlying physics, enabling the use of machine learning techniques for immediate and automatic classification \citep[e.g.][]{2008AN....329..288M,2012PASP..124.1175B,2015AJ....150...82G,2019PASP..131c8002M,2020MNRAS.tmp.1941S}. Comparably few true radio transients have been detected in untargeted searches to-date \citep{2016MNRAS.456.2321S,2018ApJ...866L..22L} and current radio transient surveys rely on manual inspection of candidates and classification using follow-up observations and archival data. Planned widefield transients searches will allow the better characterisation of transient properties, which will ultimately enable the use of automated classification algorithms.

It is naturally preferable to confirm the association between any detected radio transient as soon as possible. It is possible (albeit, unlikely) that the radio afterglow may be detected early enough that optical emission from the kilonova is still detectable, observations of which are vital in constraining merger properties like the ejecta mass. More importantly, comprehensive broadband monitoring of the non-thermal afterglow that traces the rise, peak and decline of the lightcurve, is vital in constraining merger properties.

Overall, we strongly emphasise the importance of designing follow-up strategies that are not only sufficient to detect afterglows, but also use a cadence that enables false-positives to be ruled out in a timely manner.

\subsubsection{Mergers with radio-loud hosts}
\label{subsec:radio_loud_hosts}
Resolving the afterglow from any host galaxy emission may be a decisive factor in whether the afterglow is detectable, particularly in a widefield unbiased search. GW170817 occurred in a radio-loud host galaxy and was offset by $10.31\,$arcsec \citep{2017ApJ...848L..22B,2017ApJ...848L..28L} and therefore non-standard sourcefinding techniques may have been required to find it in an unbiased search depending on the angular resolution of the data. Host galaxy offsets for short GRBs range from 0.5-75\,kpc with a median of 5\,kpc \citep{2010ApJ...708....9F,2013ApJ...776...18F,2014ARA&A..52...43B}. At 200(500)\,Mpc this corresponds to an angular offset of 10(4)\,arcsec, comparable to the angular resolution of at least some current radio telescopes. For events occuring at cosmological distances, detectable with third generation detectors (see \ref{subsec:3g_detectors}), typical offsets will be comparable to the angular resolution of the SKA. Therefore the presence of nuclear emission from the host may complicate searches, but not make them impossible. However, radio emission from star formation regions can span a much larger volume \citep{2020ApJS..248...25L} and may pose more of a problem.

By comparing the expected afterglow luminosity to the typical luminosity of both AGN and star forming galaxies, \citet{2016ApJ...831..190H} found that most mergers will not occur in galaxies that are sufficiently radio-bright to hinder radio follow-up efforts. This is true even for telescopes with angular resolution $\gtrsim 10\,$arcsec. The detectability metrics we use in this work assume that the detectability of an event is not limited by host emission.

\subsubsection{Galaxy catalogue completeness}
\label{subsec:galaxy_cat_completeness}
Most current radio telescopes lack the field of view and survey speed to carry out unbiased searches and are therefore restricted to the galaxy-targeted approach outlined in Section \ref{subsubsec:galaxy_targeting}. The effectiveness of this strategy relies upon having a complete catalogue of galaxies within the localisation volume of each event.

There is currently no all-sky galaxy catalogue that approaches completeness at the current LIGO horizon. GLADE \citep{2018MNRAS.479.2374D} is complete to a distance of $\sim 40\,$Mpc, $\sim 50$\% complete at the nominal O4 range of $170\,$Mpc and $<40$\% complete at the design range of 330\,Mpc all based on cumulative blue luminosity (which is an approximate analog for star formation, and therefore, merger rate). Additionally the galaxies in GLADE are not isotropically distributed -- the median line of sight density is $\sim 10\,\deg^{-2}$, compared to $< 1\,\deg^{-2}$ in the Galactic plane and $>10^{3}\,\deg^{-2}$ in fields covered by the HyperLEDA survey \citep{2014A&A...570A..13M}.

Other existing surveys provide more complete samples along particular lines of sight. These smaller catalogues may still be useful for follow-up of specific events -- e.g. the galaxy-targeted follow-up of GW190814 carried out by \citet{2019ApJ...884L..55G} and \citet{2020A&A...643A.113A} could have used the PRLS survey rather than GLADE. However, they do not present a general solution for follow-up of all events.

Planned all-sky surveys \citep[e.g.][among others]{2009arXiv0912.0201L,2016arXiv160607039D} will drastically improve upon existing catalogues and make it feasible to target galaxies for events at Gpc distances. However, these surveys will likely not be finished within the 5 years, by which time facilities like ASKAP and MeerKAT will be fully operational and the DSA-2000 and SKA may be coming online. These telescopes are not suited to galaxy targeting due to their wide fields of view (compared to most existing facilities), and therefore it's unlikely that future galaxy catalogues will have a tangible impact on the pointing strategies used in radio follow-up. However, they will remain extremely useful for ruling out false-positive radio transient candidates and in determining whether discovered transients are associated with host galaxies within the localisation volume.

\subsection{Follow-up of GW-identified mergers}
\label{subsec:follow_up_detectability}
To quantify prospects for detecting radio afterglows we use a simple detection metric -- the flux density at the observing frequency must exceed five times the expected thermal noise, $\sigma$. This estimate does not include the noise due to source confusion, which will not be the limiting factor for the observations discussed here.

We also compare our results to the range of gravitational wave detectors which is dependent on inclination angle, scaling as 
\begin{equation}
    \mathcal{R}(\theta_{\rm obs}) \approx 0.589\overline{\mathcal{R}}\sqrt{1+6\cos^2\theta_{\rm obs}+\cos^4\theta_{\rm obs}}
    \label{eq:gw_inc_angle}
\end{equation}
where $\mathcal{R}(\theta_{\rm obs})$ is the inclination angle dependent range and $\overline{\mathcal{R}}$ is the gravitational wave detector range found in Table \ref{tab:gw_facilities}. The variable terms in this equation are obtained by averaging eq. 3.31 of \citet{1993PhRvD..47.2198F} across the antenna pattern terms\footnote{We note that this is a crude approximation to the true survey volume, which can be more accurately calculated via a numerical Monte Carlo integral (Mandel, private communication). We have compared both methods and find the above equation underestimates the range by $\sim$10\%. However, we have chosen to use this method as it is sufficiently accurate for our purposes and is far easier to reproduce.}, while the normalisation constant is calculated by requiring $\langle \mathcal{R}(\theta_{\rm obs})\rangle = \overline{\mathcal{R}}$. \citet{2019A&A...631A..39D} obtain the same result by instead computing the gravitational wave detector horizon. While \citet{1993PhRvD..47.2198F} only calculate antenna patterns for L-shaped interferometers, the above equation is a sufficient approximation to the 3G detector configurations for our purposes as any discrepancy is negligible compared to the current uncertainty in detector specifications and sensitivity.

\begin{figure}
    \centering
    \includegraphics{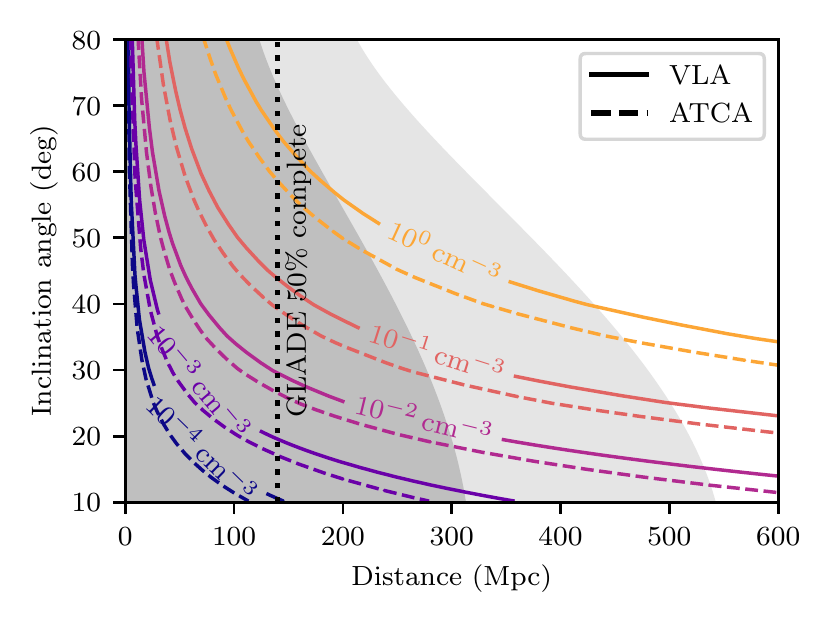}
    \caption{Maximum distance at which gravitational wave afterglows can be detected as a function of inclination angle for a range of circum-merger densities. Solid lines denote observations of 200 galaxies to a detection threshold of 75\,$\mu$Jy at 3\,GHz with the VLA, while dashed lines denote observations with the ATCA targeting 50 galaxies to a detection threshold of 70\,$\mu$Jy at 8\,GHz. The LIGO range for O4 and design specifications is shaded in dark and light grey respectively. The 50\% completeness of the GLADE catalogue is also labelled.}
    \label{fig:galaxy_targeting_range}
\end{figure}

\subsubsection{Galaxy Targeting}
Figure \ref{fig:galaxy_targeting_range} shows the maximum distance at which an afterglow can be detected for a range of circum-merger densities spanning $10^{-4}$--$1\,$\,cm$^{-3}$, using a galaxy-targeting approach with the ATCA and the VLA as outlined in Section \ref{subsec:existing_facilities}.

We note that while this strategy will allow most events to be detected at distances comparable to the LIGO range, the incompleteness of existing galaxy catalogues at these distances makes this strategy only feasible for the closest mergers ($D_L\ll 100\,$Mpc). We do not consider applying this approach to any next generation facilities as their fields of view are large enough that they contain multiple candidate hosts per pointing, and their survey speeds are large enough that an unbiased search is generally feasible.

\begin{table}
    \caption{Capabilities of unbiased searches for radio afterglows for a range of telescopes and observing strategies, including observing frequency ($\nu$), bandwidth ($\Delta\nu$), total areal coverage ($\Omega_{\rm total}$) and required observing time ($T_{\rm total}$)}
    \label{tab:untargeted_searches}
    \centering
    \begin{threeparttable}
    \begin{tabular}{lccS[table-format=1.1]S[table-format=1.1]S[table-format=1.2]S[table-format=2.1]cc}
    \hline\hline
    Telescope & $\nu$ & Strategy & $\Omega_{\rm total}$ & $S_{\rm detect}$ & $T_{\rm total}$\\
     & & {(GHz)} & {(deg$^2$)} & {($\mu$Jy)} & {(hr)}\\
    \hline
    Apertif & 1.4 & deep & 10 & 125 & 12\\
     & & wide & 40 & 250 & 12\\
    \rule{0pt}{3ex}ASKAP & 0.9 & deep & 30 & 175 & 10\\
     & & wide & 300 & 550 & 10\\
    \rule{0pt}{3ex}DSA & 1.35 & deep & 10 & 5 & 1\\
     & & wide & 100 & 5 & 2.5\\
    \rule{0pt}{3ex}MeerKAT & 1.4 & wide & 10 & 35 & 12\\
    \rule{0pt}{3ex}ngVLA & 2.4 & wide & 10 & 5 & 10\\
     & & ultra-wide & 100 & 25 & 10\\
    \rule{0pt}{3ex}SKA-1 & 1.43 & wide & 10 & 10 & 10\\
     & & ultra-wide & 100 & 40 & 10\\
    \rule{0pt}{3ex}SKA-2 & 1.43 & wide & 10 & 1 & 10\\
     & & ultra-wide & 100 & 4 & 10\\
    \rule{0pt}{3ex}VLA & 1.5 & wide & 5 & 75 & 12\\
    \hline\hline
    \end{tabular}
    \end{threeparttable}
\end{table}

\begin{figure*}
    \centering
    \includegraphics{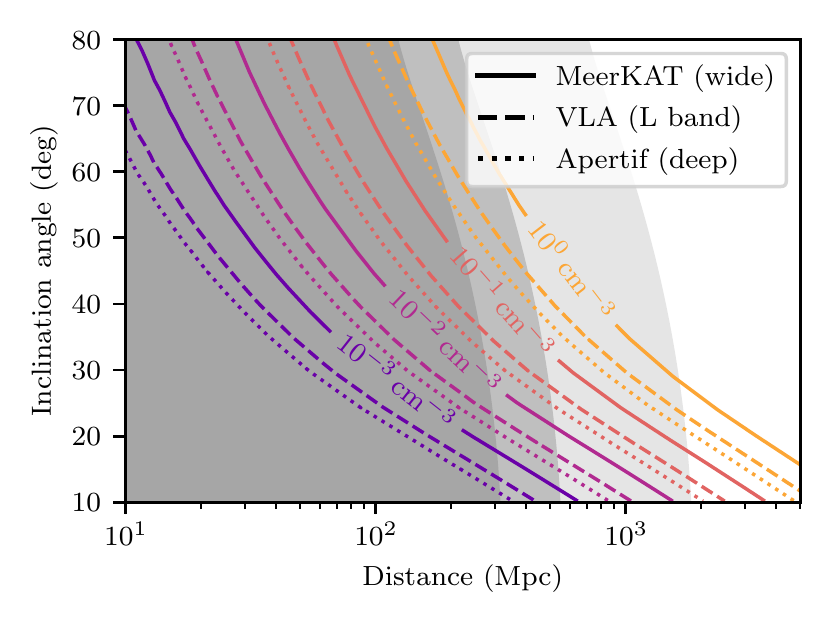}
    \includegraphics{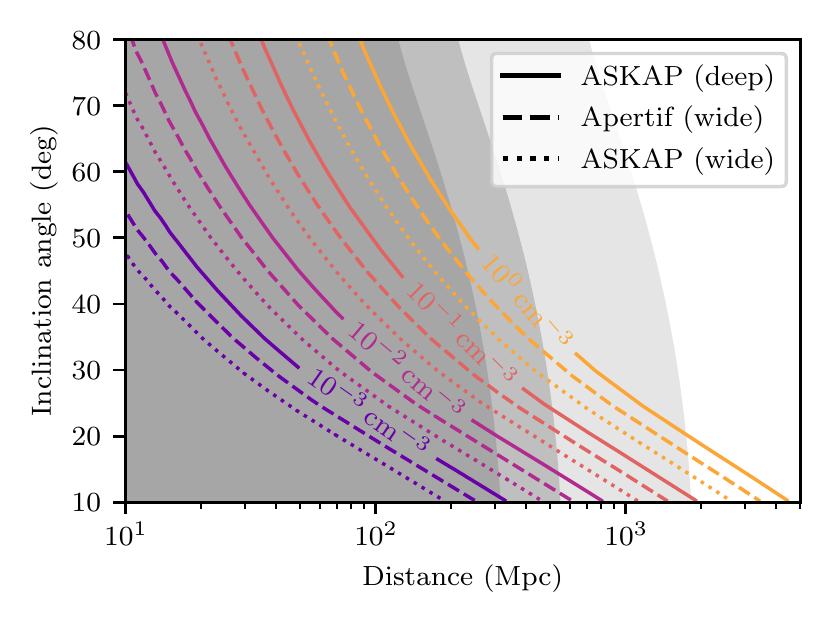}
    \caption{Similar to Figure \ref{fig:galaxy_targeting_range}. Left: unbiased observations of events localised to $\leq 10\,\deg^2$ with MeerKAT, Apertif and the VLA. Right: unbiased observations with Apertif (covering $40\,\deg^2$) and ASKAP (deep covering $30\,\deg^2$ and wide covering $300\,\deg^2$). The gravitational wave detector range for O4, A+ and Voyager are shown in increasingly light tones of grey.}
    \label{fig:current_unbiased}
\end{figure*}

\begin{figure*}
    \centering
    \includegraphics{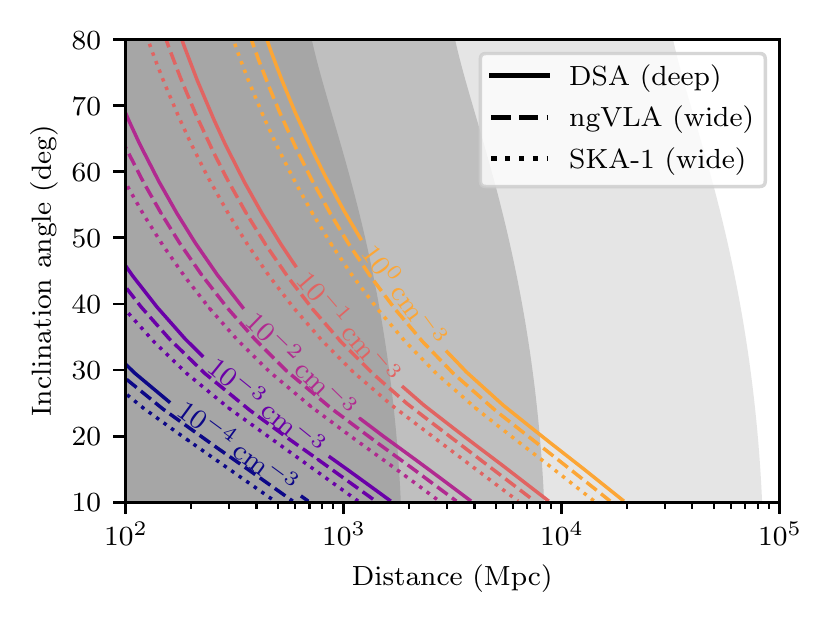}
    \includegraphics{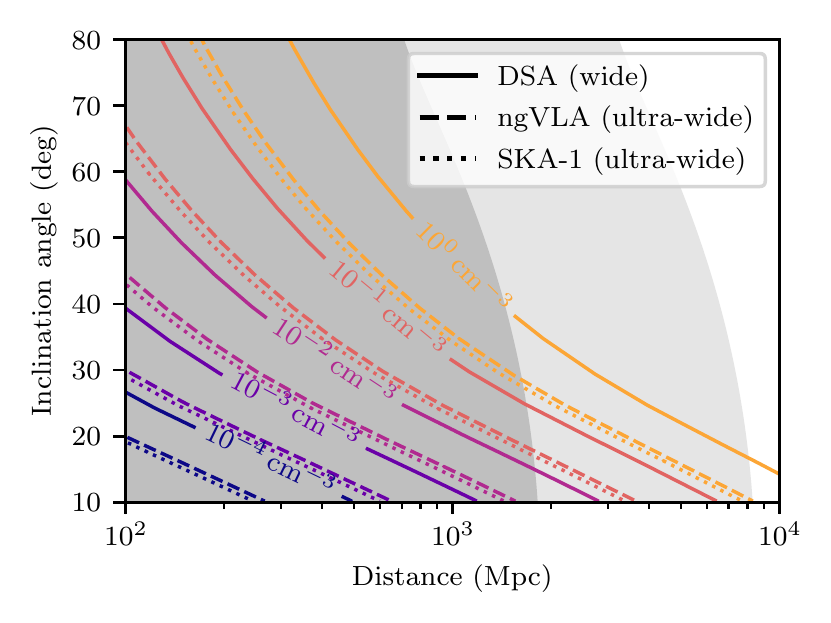}
    \caption{Similar to Figure \ref{fig:galaxy_targeting_range}. Unbiased observations of events localised to $\leq 10\,\deg^2$ (left) and $\leq 100\,\deg^2$ (right) with the SKA-1, DSA-2000 and ngVLA. The range of a nominal SKA-2 design is $\sim 3$ times further than the SKA-1. The detector range of Voyager, a preliminary 3G detector with a 5\,Gpc range, and 3G detectors are shown in increasingly light tones of grey.}
    \label{fig:future_unbiased}
\end{figure*}

\subsubsection{Unbiased Searches}
We therefore turn our focus to the unbiased searches described in Table \ref{tab:untargeted_searches}, which we split into four broad categories.

Figure \ref{fig:current_unbiased} shows the detectability of events in unbiased searches with current facilities. Most on-axis mergers, as well as most off-axis mergers occuring in dense environments, localised to $\leq 10\,\deg^2$ detected with current gravitational wave facilities and the A+ upgrade will be detectable with MeerKAT and some will be detectable with the VLA and Apertif. However, we note that only a small fraction of events will be localised this well with these detectors. LIGO Voyager will have better localisation capabilities, and some events will produce afterglows that are detectable out to the detector horizon.

For events that are localised to tens of square degrees, comparable to the median localisation for 2G detectors \citep{2020LRR....23....3A}, we consider follow-up with ASKAP and Apertif. Figure \ref{fig:current_unbiased} also shows the maximum distance at which afterglows will be detected, and we find that it is feasible to detect the afterglow produced by most on-axis mergers with current gravitational wave detectors.

Figure \ref{fig:future_unbiased} shows the same metrics applied to the DSA-2000, ngVLA and SKA-1 compared to the range of Voyager and 3G detectors for events localised to $\leq 10\,\deg^2$ and $\leq 100\,\deg^2$. We find that while the majority of events detected with Voyager will be accompanied by detectable afterglows, it will not be possible to detect afterglows in widefield follow-up of the most distant events discovered by a complete 3G network. However, we note that the median localisation achievable with a complete 3G network is $\sim 1\,\deg^2$. Therefore widefield searches will not be necessary for most events, and the targeted single-pointing strategy outlined in Section \ref{subsec:monitoring_known_counterparts} may be a more useful metric. 

\begin{figure}
    \centering
    \includegraphics{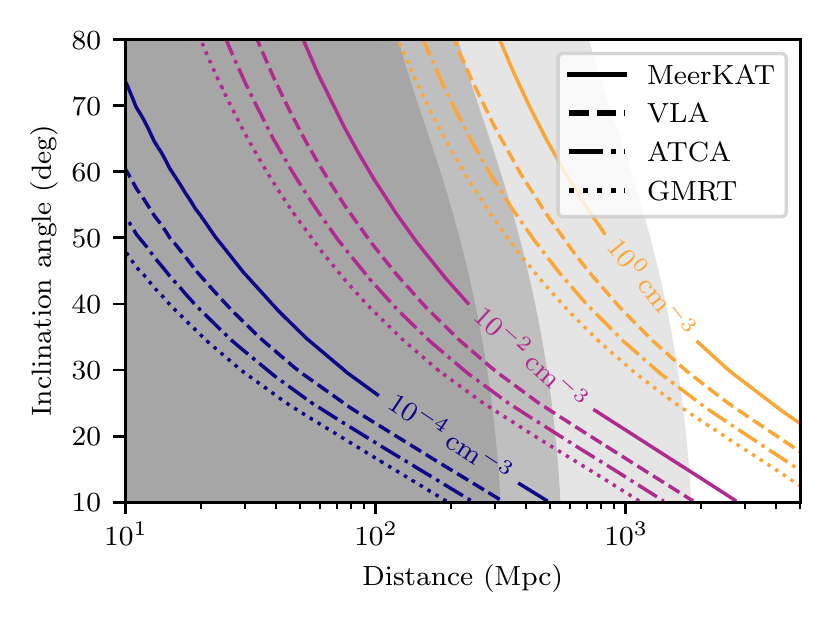}
    \caption{Similar to Figure \ref{fig:galaxy_targeting_range} showing targeted single-pointing observations with existing radio telescopes. The range of gravitational wave detectors for O4, A+ and Voyager are shown in increasingly light tones of grey.}
    \label{fig:current_monitoring}
\end{figure}

\begin{figure}
    \centering
    \includegraphics{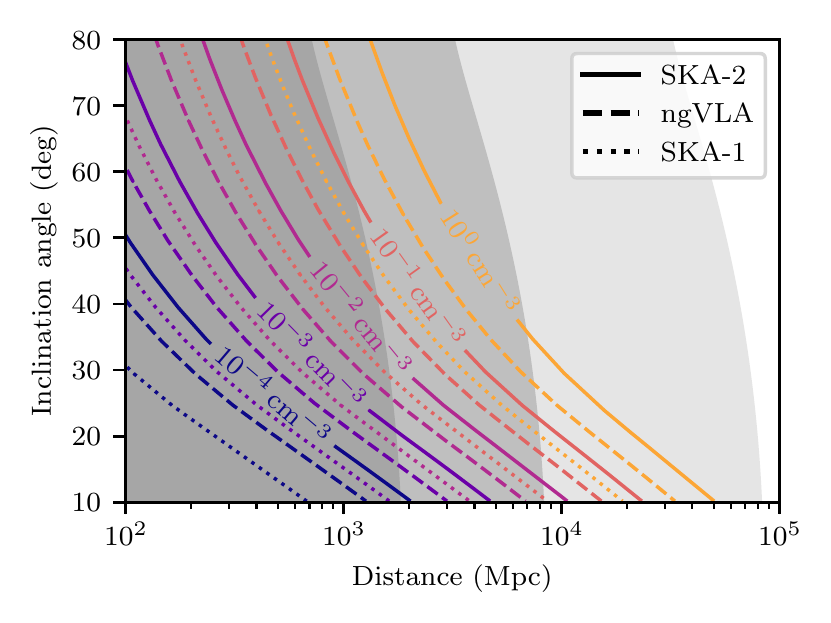}
    \caption{Similar to Figure \ref{fig:galaxy_targeting_range} showing targeted single-pointing observations with planned radio telescopes. The range of the DSA-2000 is comparable to the range of the SKA-1. The range of Voyager, a preliminary 3G detector with a 5\,Gpc range, and 3G detectors are shown in increasingly light tones of grey.}
    \label{fig:future_monitoring}
\end{figure}

\subsubsection{Monitoring electromagnetic  counterparts}
\label{subsec:monitoring_known_counterparts}
Figure \ref{fig:current_monitoring} shows prospects for detecting radio emission with current radio facilities from events that have been localised by the detection of an electromagnetic counterpart. Assuming that neutron star mergers occur in comparably dense environments to short GRBs \citep[$n \sim 10^{-2}\,{\rm cm}^{-3}$][]{2015ApJ...815..102F} we find that most neutron star mergers detected during O4, and a large fraction with the A+ configuration, should produce radio emission that is detectable with deep single pointing observations.
However current facilities will not be sufficient for a comprehensive census of radio afterglows as we move towards the 3G era -- only on-axis mergers occuring in dense environments ($n\gtrsim 10^{-1}\,{\rm cm}^{-3}$) will be detectable at the Voyager horizon.

The sensitivity of future radio telescopes will partially address this problem. Figure \ref{fig:future_monitoring} shows the detectability horizon for the ngVLA and both phases of the SKA (with the DSA-2000 horizon comparable to that of the SKA-1). While most events detected with Voyager will be within range of future radio telescopes, the most distant events detected with 3G detectors will be well beyond the range of even the SKA-2. However, these detectors will detect thousands of events per year and therefore the limiting factor in obtaining a census of radio afterglows will be the amount of telescope time available rather than the current scenario which is limited by a lack of events.

\begin{figure*}
    \centering
    \includegraphics{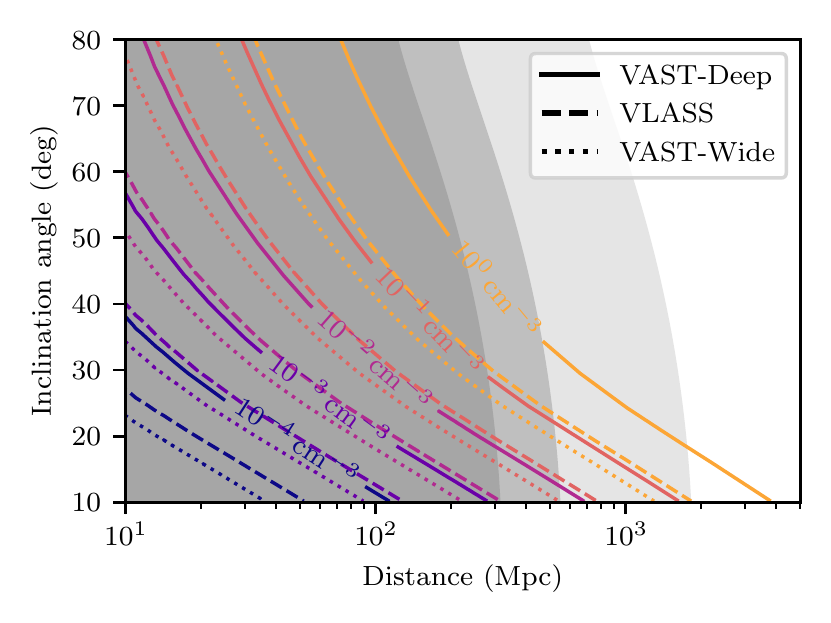}
    \includegraphics{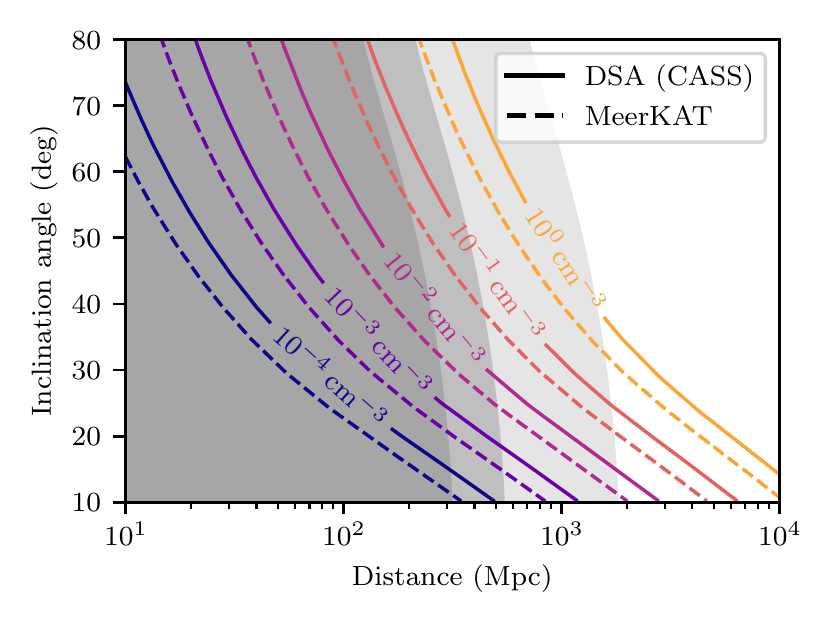}
    \caption{Detectability of radio afterglows for planned (left) and potential future (right) untargeted transients surveys. The detector range for O4, A+ and Voyager are shown in increasingly light shades of grey.}
    \label{fig:transients_surveys_serendipitous}
\end{figure*}

\begin{figure}
    \centering
    \includegraphics{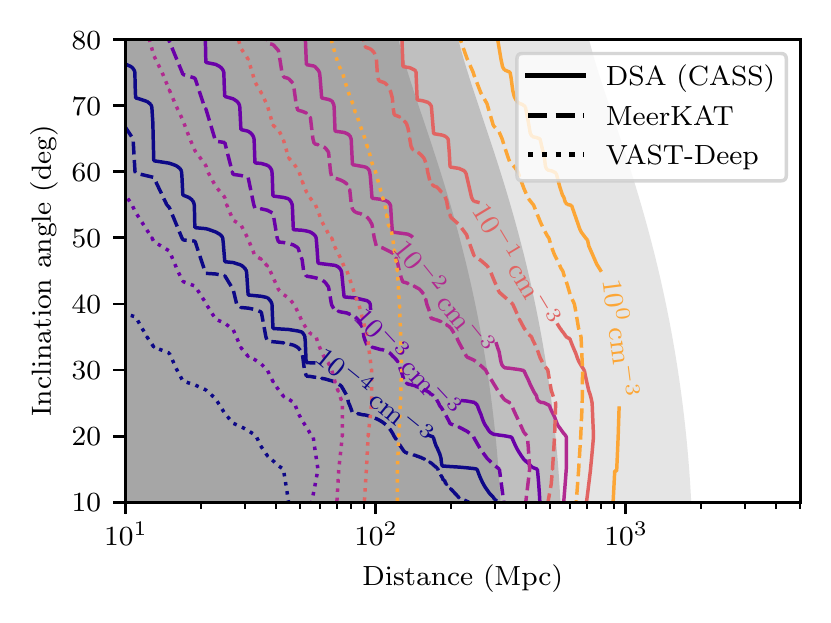}
    \caption{Detectability of a complete sample of radio afterglows (i.e. where the signal remains detectable for longer than the survey cadence) in planned and potential future untargered transients surveys. The detector range for O4, A+ and Voyager are shown in increasingly light shades of grey. As the inclination angle of the merger decreases the detectability becomes limited by the survey cadence rather than sensitivity. This effect is more pronounced for mergers occurring in denser environments.}
    \label{fig:transients_surveys_full}
\end{figure}

\begin{table}
    \caption{Ongoing, upcoming and idealised transients surveys. The fraction of the total sky covered by the survey is in column 3, while $S_{\rm det}$ corresponds to a $5\sigma$ detection threshold based on expected image noise. The MeerKAT survey is a theoretical idealised survey and there are no current plans to undertake it.}
    \label{tab:transient_surveys}
    \centering
    \begin{tabular}{lccccc}
    \hline\hline
    Survey & $\nu$ & Sky coverage & Cadence & $S_{\rm det}$\\
     & (GHz) & & (months) & ($\mu$Jy)\\
    \hline
    VLASS & 3 & 0.82 & 32 & 600\\
    \rule{0pt}{4ex}VAST-Wide & 0.9 & 0.23 & daily & 2500\\
    VAST-Deep & 0.9 & 0.23 & 8 & 250\\
    \rule{0pt}{4ex}MeerKAT & 1.4 & 0.12 & 4 & 20\\
    DSA (CASS) & 1.35 & 0.75 & 4 & 10\\
    \hline\hline
    \end{tabular}
\end{table}
\subsection{Serendipitous detections and orphan afterglows}
We also consider the detection of afterglows from known gravitational wave events and orphan afterglows (i.e. events with no previous EM/GW detection), in the transients surveys outlined in \ref{subsec:serendipitous}. Table \ref{tab:transient_surveys} shows the properties of the surveys, and Figure \ref{fig:transients_surveys_serendipitous} shows the application of the same detectability metrics as above. 

To determine the capability of surveys to obtain a complete sample of mergers occuring within their footprint, we also apply an additional constraint of the afterglow remaining detectable for a time corresponding to the survey cadence. This ensures that the afterglow will be detected in at least one epoch of the survey, although we note that multiple detections and multi-wavelength follow-up will be required to confirm the detection and classify it. The results of this change are shown in Figure \ref{fig:transients_surveys_full}, where we have excluded VLASS (as its slow cadence results in effectively zero range) and VAST-Wide (as its daily cadence results in no significant changes to the result above). Sensitivity remains the dominant limiting factor for off-axis events, while on-axis events occurring in denser environments are limited by survey cadence as their emission peaks at earlier times with a shorter turnover period. We note that these results cannot be scaled using equation \ref{eq:flux_scaling}, as the rise and decay of the lightcurve are also dependent on our assumptions of $E_{\rm iso}$ and $\Gamma(\theta)$.

While planned transients searches will not provide complete samples of afterglows due to their limited sky coverage, these results demonstrate that it is worthwhile carrying out targeted searches for afterglows within those datasets. These surveys may also make it feasible to search for counterparts to poorly localised events that do not individually warrant follow-up observations. We also note that once Voyager begins operations very few orphan afterglows will be detected, as only on-axis events occuring in the most dense environments will be detectable at distances beyond the Voyager horizon, although this does not consider the duty cycle of the detector network.

\section{Conclusions}

In this work we have summarised the synergies between radio telescopes and gravitational wave detectors, encompassing both existing facilities and planned or proposed facilities spanning the coming decades. We demonstrate that while targeting potential host galaxies proved useful in the follow-up of GW170817, this method will be less feasible in future follow-up due to limited catalogue completeness at distances comparable to the range of gravitational wave detectors. Additionally, the larger field of view of future telescopes is more conducive to unbiased widefield searches that target the localisation region of the merger. We find that these searches with current facilities will be capable of detecting mergers at hundreds of Mpc, while future facilities will be able to detect mergers at Gpc distances. Widefield transients surveys will provide serendipitous coverage of events and may also detect afterglows of events beyond the detector horizon, and those that occur during detector downtime. 

Radio observations can be used to place constraints on properties of the merger outflow and the circum-merger environment, and for events with counterparts detected at other wavelengths, we find that current radio facilities are capable of detecting some afterglows at the Voyager horizon, while future facilities will detect afterglows at distances up to tens of Gpc for the most on-axis events. However, lightcurve monitoring alone is insufficient to completely constrain the geometry of the merger, and we also discuss possible ways of breaking model degeneracies.

\section*{Acknowledgements}
We thank Rob Fender, Samaya Nissanke, Andrew Zic, and Ilya Mandel for useful discussions. DD was supported by an Australian Government Research Training Program Scholarship. TM acknowledges the support of the Australian Research Council through grant DP190100561. DLK was supported by NSF grant AST-1816492. Parts of this research were conducted by the Australian Research Council Centre of Excellence for Gravitational Wave Discovery (OzGrav), project number CE170100004. This research has made use of NASA's Astrophysics Data System Bibliographic Services and the FRB Theory wiki \citep{2019PhR...821....1P}.

\textit{Software:} Astropy \citep{2018AJ....156..123A}, Jupyter \citep{Kluyver:2016aa}, Matplotlib \citep{2007CSE.....9...90H}, NumPy \citep{harris2020array}, SciPy \citep{Virtanen_2020}

\section*{Data availability}
The data underlying this article will be shared on reasonable request to the corresponding author.


\bsp
\label{lastpage}
\end{document}